\documentclass[conference]{IEEEtran} \IEEEoverridecommandlockouts
\usepackage{cite}
\usepackage{amsmath,amssymb,amsfonts}
\usepackage{algorithmic}
\usepackage{graphicx}
\usepackage{textcomp}
\usepackage{xcolor}
\usepackage[bottom]{footmisc}
\usepackage[shortlabels]{enumitem}
\usepackage{mathtools}
\usepackage{url}%hyperref
\usepackage{svg}
\usepackage{todonotes}
\usepackage{caption}
\usepackage{url}
\usepackage{multirow}
\usepackage{subcaption}
\usepackage{fancyhdr}
\usepackage[absolute,overlay]{textpos}

\def\BibTeX{{\rm B\kern-.05em{\sc i\kern-.025em b}\kern-.08em
    T\kern-.1667em\lower.7ex\hbox{E}\kern-.125emX}}
\begin{document}

\fancyhf{}
\renewcommand{\headrulewidth}{0pt}
\fancyfoot[c]{}
\fancypagestyle{FirstPage}{
%\lfoot{978-1-6654-8215-8/22/\$31.00 \copyright2022 IEEE} 
}

%%%%%%%% Paper title 
\title{Towards Intuitive HMI for UAV Control}
%\footnotesize

%\author{Filip Zorić$^{1}$ \quad Goran Vasiljević$^{1}$ \quad Zdenko Kovačić$^{1}$
%\IEEEauthorblockA{\textit{University of Zagreb, Faculty of Electrical Engineering and Computing} \\
%\textit{Laboratory for Autonomous Robotics and Intelligent Control Systems (LARICS)}\\
%Zagreb, Croatia\\
%filip.zoric@fer.hr}

% Provjeriti li imaju li zarezi
\author{
\IEEEauthorblockN{Filip Zorić}
\IEEEauthorblockA{\textit{Faculty of EE\&C} \\
\textit{University of Zagreb}\\
Zagreb, Croatia \\
filip.zoric@fer.hr}
\and
\IEEEauthorblockN{Goran Vasiljević}
\IEEEauthorblockA{\textit{Faculty of EE\&C} \\
\textit{University of Zagreb}\\
Zagreb, Croatia \\
goran.vasiljevic@fer.hr}
\and
\IEEEauthorblockN{Matko Orsag}
\IEEEauthorblockA{\textit{Faculty of EE\&C} \\
\textit{University of Zagreb}\\
Zagreb, Croatia \\
matko.orsag@fer.hr}
\and
\IEEEauthorblockN{Zdenko Kovačić}
\IEEEauthorblockA{\textit{Faculty of EE\&C} \\
\textit{University of Zagreb}\\
Zagreb, Croatia \\
zdenko.kovacic@fer.hr}
}

\maketitle

%\thanks{}
%\thanks{$^{1}$ Filip Zorić, Goran Vasiljević and Zdenko Kovačić are with University of Zagreb, Faculty of Electrical Engineering and Computing, Laboratory for Robotics and Intelligent Control Systems (LARICS), Unska 3, Zagreb 10000, Croatia}}

% Dodati ovaj dio za EU: 
% 978-1-6654-8215-8/22/$31.00 ©2022 European Union

%% Dodati po potrebi još nekoga 

\maxdeadcycles=20000
\begin{textblock*}{14.9cm}(3.2cm,0.75cm) %
	{\footnotesize © 2022 IEEE.  Personal use of this material is permitted.  Permission from IEEE must be obtained for all other uses, in any current or future media, including reprinting/republishing this material for advertising or promotional purposes, creating new collective works, for resale or redistribution to servers or lists, or reuse of any copyrighted component of this work in other works.}
\end{textblock*}

\begin{abstract}

In the last decade, UAVs have become a widely used technology. As they are used by both professionals and amateurs, there is a need to explore different control modalities to make control intuitive and easier, especially for new users. In this work, we compared the most widely used joystick control with a custom human pose control. We used human pose estimation and arm movements to send UAV commands in the same way that operators use their fingers to send joystick commands. Experiments were conducted in a simulation environment with first-person visual feedback. Participants had to traverse the same maze with joystick and human pose control. Participants' subjective experience was assessed using the raw NASA Task Load Index. 
%TODO: Write abstract 
% Check how to write good abstract? 
\end{abstract}


% All IEEE keywords: https://www.ieee-ras.org/publications/toh/keywords
% TODO: Add keywords -> robotic manipulation 

%%%%% Sections 
\thispagestyle{FirstPage}
\section{Introduction}
\label{section:Introduction}

%% OK for starters 
In recent years, number of active registrations of unmanned aerial vehicles (UAVs) has increased ten-fold which implies they are entering mainstream.\footnote{\url{https://www.faa.gov/data_research/aviation/aerospace_forecasts/media/unmanned_aircraft_systems.pdf}}
As the technology moves into the mainstream, the number of use cases and opportunities is also growing by leaps and bounds. In addition to technological advancements in UAV control techniques and flight dynamics, there have also been significant advances in modes of operation. Most notably, the commercial availability of first-person view (FPV)\footnote{\url{https://www.dji.com/au/newsroom/news/dji-reinvents-the-drone-flying-experience-with-the-dji-fpv}} systems has paved the way for hobbyists to perform flight maneuvers and capture video footages which were exclusively in domain of licensed aerial cinematographers just few years ago. 

Currently, UAVs are typically used for surveillance\cite{Agbeyangi2016}, monitoring\cite{Ren2019},\cite{Butila2022}, and inspection tasks in a variety of industries\cite{Li2021},\cite{Jordan2018}. They are also used for entertainment, communication, sports, law enforcement and construction\cite{Herdel2022}. To a lesser extent, UAVs are used for transportation purposes such as parcel delivery and passenger transportation\cite{Kellerman2020}. Some of the interesting applications are precision agriculture, inspection of pipelines and wind turbines, monitoring of facilities or power lines \cite{Kratky2021}. Most of these applications require a trained operator to control the UAV during the task. Human-machine interface (HMI) refers to any technology that enables human-machine interaction.

% Not necessary 

% Footnote without mark conference.tex part

In the literature, we find terms such as Human-Drone Interaction (HDI)\cite{Funk2018}, Human-Robot Interface/Interaction (HRI)\cite{Stiefelhagen2007}, and HMI used interchangeably. The term HMI is the most appropriate for our purpose and will be used in the remainder of this paper.

The interaction between human and UAV is divided into two types of control, namely non-gestural and gestural\cite{Cherpillod2019}. Non-gestural control is based on joystick, RC controllers, touch screens, brain-machine interaction based on electroencephalographic (EEG) signals, and gaze tracking control. Gestural control uses gesture recognition techniques to control the UAV. Gestural control can use upper body gestures, hand gestures as well as head movements. Gestural control is divided into first-person view and third-person view. In the first-person view operator sees the world from UAV point of view, while in the third-person view operator observes the UAV as long as it is in his line of sight. \textit{In our case, we developed continuous gestural control based on human pose estimation in first-person view.}

\subsection{Problem statement}

In this paper, we are exploring how different control modalities of human machine interface affect UAV control. We compare conventional joystick control with developed human pose control. We use raw NASA Task Load Index\cite{NASAtlx} to assess workload difference between control modalities. 

\subsection{Contributions}

The motivation to use human pose estimation for continuous UAV control, stems from the wide range of aerial manipulation tasks that could benefit from such an easy-to-deploy and intuitive control system. In particular, if the operator could use the same control modality for both the UAV and the robotic manipulator. We wanted to show that it is possible to use existing technologies to create an intuitive yet efficient control modality without having to equip the operator with numerous electronic sensors. We also developed experimental procedure to evaluate such control modality and compare it to existing control modality with multiple user study. 

The contributions of this paper are: 
\begin{enumerate}
    \item Continuous gestural UAV control based on human pose estimation with first person view.
    \item Evaluation of the developed control method using experimental procedure, raw NASA TLX and qualitative analysis.
\end{enumerate}

%% Contributions 
%%%%%%%%%%%%%%%%%%%%%%%%
% To the best of our knowledge this is first paper that uses human pose estimation as control modality paired up with visual feedback. 

% First continuous human pose control based on human pose estimation, keypoints used for continuous control (discrete control based on gestures?) 

%%%%%%%%%%%%%%%%%%%%%%%%

% How to write we will use HMI and why it fits our purpose best? 
% Maybe explain which term fits our purpose: 
% 1) HRI --> to much oriented on interaction (which is from my perspective action, reaction, in our use-case, reaction is expected, it's not https://www.overleaf.com/project/61fbb3d3e55bde23301519c5
% unexpected 
% 2) HDI --> drones are too general, although they're mostly used for UAVs, all remotely controlled systems can be drones You: a imas i ovo dosta impresivno
% https://blogs.scientificamerican.com/guest-blog/what-is-a-drone-anyway/
% 3) HMI --> human machine interface is good enough because in this case we're constrainted to UAV, but we could basically use same interface to operate different machine (arm) 

% Used info
% --------------------- 
% 1) https://botlink.com/blog/whats-the-difference-between-a-drone-uav-and-uas
% 2) https://blogs.scientificamerican.com/guest-blog/what-is-a-drone-anyway/

% In this paper we present novel method of UAV control based on gestures as well as comparsion of gesture control and remote controlled. 

% https://helvetis.com/uav-wind-turbine-blades-inspection-drone/

% Gorrila arm syndrome 
% Could be cool in introduction 
% Add maybe info about number of registered UAVs in USA (FAA data) ?

\subsection{Paper organization}
The paper is divided into seven sections. The second section contains related work and gives a brief overview of similar work in the field. The third section contains a formalization and explanation of human pose control. The fourth section describes the experimental setup. The fifth section contains the experimental results. The sixth section contains conclusions and plans for future work. The seventh section contains an appendix listing all additional information.

\section{Related work}
\label{section:Related work}

% Add quotation to Pose2Drone and OpenPose
Recent available work in the area of HMIs for UAVs based on gesture recognition can be found in \cite{Marinov2021}. Besides gesture recognition from human pose, distance and orientation of operator are also used as control modality. Paper presents several software modules, namely: RGB module, human pose estimation module, gesture recognition module, and depth estimation module. For the human pose estimation module, OpenPose\cite{Cao2018} and logical rules based on the Euclidean distance between the detected joints are used to perform view estimation and gesture classification. The view estimation is the orientation of the user with respect to the camera and the gesture classification is the classification into one of the eleven predefined gestures which are used to send commands. 
% Our work differs from Pose2Drone in context of using pure information from human pose estimation for sending commands 

In survey paper about HDI \cite{Tezza2019}, Tezza et al. categorized human-drone interactions and presented different roles human can have in such interactions. Categories regarding interaction are: control modalities, human-drone communication, proxemics and novel use-cases. Control modalities can be speech, gesture, brain-computer interface. Human-drone communication explores directionality of communication as well as intent. Proxemics refers to certain functional characteristics such as interaction distance and comfortable approach distance. 
Roles that human can take in human-drone interaction are as active controller, recipient, social companion and supervisor. In our case, control modality is gesture and user is active controller. 
% Maybe add where our paper sits in respect to that

% Natural controller -> maybe term it 
Cherpillod et al. \cite{Cherpillod2019} showed how different operators fly a fixed-wing UAV through waypoints using a platform with haptic, visual (FPV), and vestibular feedback. The motion of the operator's hands is converted into UAV motion, and head motion is converted into gimbal motion. The authors explain that the ultimate goal is more natural and effective control of the robot, which is indeed proven by the various quantitative and qualitative measurements made during the experiments.

Fernandez et al. \cite{Fernandez2016} presented natural user interface for multi-modal human-drone interaction. Different modalities for control which have been tested with unprofessional operators, show promising results especially for people with disabilities. The paper also includes a brief overview of user interface development, from command line interfaces (CLI) to graphical user interfaces (GUI) to natural user interfaces (NUI) and the key elements used in such interfaces. Two main media used to implement reliable HDI are speech-based and gesture based NUIs. Our work can be categorized as gesture based NUI. 

Macchini et al. \cite{Macchini2020} explored relation of users spontaneous motion with designing optimal motion-based HRIs. They concluded that spontaneous motion is not enough to design optimal motion-based HRI. They also compared motion-control HRI using torso, arm and hand, and found out that most of the participants found hand as most suitable and intuitive for motion-control of UAV. 

Zhao et al. \cite{Zhao} hand gestures and gamepad interface are compared for locomotion in virtual environments. Four different hand gestures are compared through two virtual locomotion tasks. Designed gestures are intuitive and some of the ideas could be extended to human pose estimation. Tasks are divided into target pursuit and waypoint following. Experimental evaluation of hand gestures, as well as used metrics could be used as good basis for performing further analyses of HMIs for any kind of teleoperation.       

Pfeil et al. \cite{Pfeil2013} explored 3D gesture metaphors for interaction with Unmanned Aerial Vehicles. Presented 3D gesture metaphors are: first person, game controller, the throne, proxy manipulation and seated proxy. Gesture metaphors are evaluated qualitatively and quantitatively in user studies. Game controller metaphor is quite similar to our human pose control, however it is based on Kinect pose estimation and logic is a bit different because it incorporates depth. Proxy control refers to synchronized movement of both arms that imitate UAV movement. For moving forward, participants had to push both arms forward. For roll, participants had to lower one arm in reference to another arm to achieve roll in that direction. Although authors expected game controller metaphor to be preferred by participants, results were different. Proxy control was most prefered by participants to control UAV. Metaphors that enabled decoupled control with both arms, such as game controller, made participants to preform worse in comparison to metaphors were control was coupled to both arms. 

Miehlbradt et al. \cite{Miehlbradt2018} presented body machine interface (BoMi) for the accurate control of UAVs. Identification methodology was developed that utilizes body-machine patterns for developing more accurate and intuitive teloperation systems. Authors also reported that users using only torso for UAV control outperformed those that used torso and arms.  

PinpointFly \cite{Chen2019} developed by Chen et al. is an egocentric drone interface that allows pilots to arbitrarily position and rotate a drone using position-control direct interactions on a see-through mobile AR with a virtual cast shadow. To verify interface, authors conducted two user studies with System Usability Scale (SUS) and Nasa Task Load Index (NASA-TLX). 

% Ovaj dio mozda izbaciti? 
%To best of our knowledge, most of the papers in the field of UAV control based on gestures use Microsoft Kinect SDK for acquiring gestures \cite{Gio2021}, \cite{Sanna2013} or some low-cost camera sensor and classical image processing algorithms \cite{Sun2017}. 

% Add paper of spontanous movement lead to better HMI design 
% Notes at home 

\section{Human pose control}
\label{section:Human pose control}

In this section we present human pose control. The human pose control is divided into: human pose estimation, zone definition, reference generation, and visual feedback. Human pose estimation is used to determine the current 2D pose of a person from the image. Zone definition refers to the image zones used for control. The reference generation subsection explains how UAV references are generated from human pose estimation and control zones. Visual Feedback introduces all the elements used for visual feedback.

% What to write here? 

% Human pose estimation 
%------------------------
% Zone definition 
% - how i defined zones 
%------------------------
% Reference generation 
% - conditions that generate reference based on pixel position 
%------------------------
% Image transformation 
% - mirroring --> Check where mirroring is used
%------------------------

\subsection{Human pose estimation}

For human pose estimation, Simple Baselines \cite{xiao2018simple} model included in ROS wrapper was used. Output of human pose estimation is $\mathbf{h_p} \in \mathbb{R}^{16 \times 2}$. Each row in $\mathbf{h_p}$ corresponds to normalized pixel position $(p_x, p_y)$ in the image plane.
\begin{equation}
    p_x, p_y \in [ 0, 1]    
\end{equation}

Key points correspond to different joint positions as \textit{head, neck, shoulders, elbows, hand ankles, hips, knees, leg ankles}. We use position of right and left hand ankles in image plane as shown in equations \eqref{eq:hand_r} and \eqref{eq:hand_l}. As network backbone ResNet-152 trained on MPII\cite{andriluka14cvpr} dataset was used. 
Pixels of hand ankles are defined as: 
\begin{equation}
    \mathbf{h_{p10, :}} = (p_{rx}, p_{ry})
    \label{eq:hand_r}
\end{equation}
\begin{equation}
    \mathbf{h_{p15, :}} = (p_{lx}, p_{ly})
    \label{eq:hand_l}
\end{equation}

To prevent unwanted reference jumps, the averaging filter was added to human pose estimation:
%TODO: Fix step size as param 
\begin{equation}
        \mathbf{\widetilde{h}_p}[k] = \frac{1}{n} \sum_{j=0}^{n-1} \mathbf{h_p}[k-j]
\end{equation}

\subsection{Zone definition}

\begin{figure}[h]
\centering
\includegraphics[width=1.0\columnwidth]{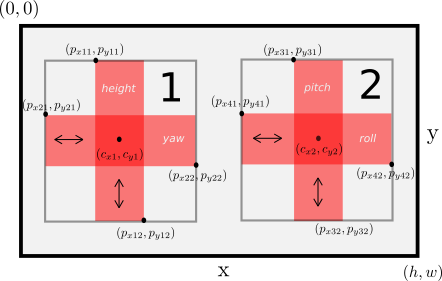}
\caption{Zones placement on image plane which are used as visual feedback for operators who generate references by arm movement}
\label{fig:zones}
\end{figure}

There are two main zones of interest, one for each arm. Moving hand ankles in zones is scaled into references sent to UAV. It is possible to control attitude and height of UAV.  Zone one is used to send height and yaw references, and zone two is used to send pitch and roll references. Zones are defined as shown in Figure \ref{fig:zones}.

% TODO: Reformulate this part 
 Control zones are defined as paired tuple of pixels that represent upper left corner and bottom right corner of the zone as follows: 
\begin{equation}
    c_{z1} = [(p_{x21}, p_{y11}), (p_{x22}, p_{y12})]
    \label{eq:cz1}
\end{equation}
 \begin{equation}
    c_{z2} = [(p_{x41}, p_{y31}), (p_{x42}, p_{y32})]
    \label{eq:cz2}
 \end{equation}
 
 Vertical motion in zone one is used for the height control. Horizontal motion in zone one is used for the yaw angle control. Horizontal motion in zone two is used for the roll control. Vertical motion in zone two is used for the pitch control.

\subsection{Reference generation}

Dead zones are defined as follows: 
\begin{equation}
    d_1 = [(p_{x11}, p_{y21}), (p_{x12}, p_{y22})]
    \label{eq:dz2}
\end{equation}
\begin{equation}
    d_2 = [(p_{x31}, p_{y41}), (p_{x32}, p_{y42})]
    \label{eq:dz1}
\end{equation}

Reference generation starts when $\tilde{h}_{10, :}$ and $\tilde{h}_{15, :}$ are located inside of the dead zones. When operator arms are in dead zones, $r_i = 0$. The references $r_i$ shown in \eqref{eq:ref} are used to send attitude and altitude references to the UAV as shown in equations \eqref{eq:theta}, \eqref{eq:phi}, \eqref{eq:psi}, \eqref{eq:z}. 

\begin{equation}
    r = (r_1, r_2, r_3, r_4), \quad r_{i}\in [-1, 1]
    \label{eq:ref}
\end{equation}

% Check for sets
The averaged hand ankle position $\widetilde{p}$ is used to generate the reference proportional to the distance from the zone center when placed within the zone as follows: 
\begin{equation}
  r_1 =
    \begin{cases}
      \frac{c_{y1} - \widetilde{p}_{ly}}{(p_{y12} - p_{y11})/2} &   \text{if}\  \widetilde{p}_{ly} \in (p_{y11}, p_{y21}) \cup (p_{y22}, p_{y12}) \\
      0 & \text{otherwise}
    \end{cases}   
    \label{eq:r1}
\end{equation}

\begin{equation}
r_{2} =
    \begin{cases}
    \frac{c_{x1} - \widetilde{p}_{lx}}{(p_{x22} - p_{x21})/2} & \text{if}\ \widetilde{p}_{lx} \in (p_{x21}, p_{x11}) \cup (p_{x12}, p_{x22}) \\
    0 & \text{otherwise}
    \end{cases}
    \label{eq:r2}
\end{equation}

\begin{equation}
r_{3} =
    \begin{cases}
    \frac{c_{y2} - \widetilde{p}_{ry}}{(p_{y32} - p_{y31})/2} & \text{if}\ \widetilde{p}_{ry} \in (p_{y31}, p_{y41}) \cup (p_{y42}, p_{y32}) \\
    0 & \text{otherwise}
    \end{cases}
\end{equation}

\begin{equation}
r_{4} =
    \begin{cases}
    \frac{c_{x2} - \widetilde{p}_{rx}}{(p_{x42} - p_{x41})/2} & \text{if}\ \widetilde{p}_{rx} \in (p_{x41}, p_{x31}) \cup (p_{x32}, p_{y42}) \\
    0 & \text{otherwise}
    \end{cases}
\end{equation}

\subsection{Visual feedback}

Main component of the visual feedback is camera stream from the UAV. Other components that are also included in visual feedback are following: 
\begin{enumerate}
    \item \textbf{compass} informs operator about current UAV yaw \label{element:1}
    \item \textbf{artifical horizon} informs operator which is current UAV attitude regarding roll and pitch angles \label{element:2}
    \item \textbf{human pose control feedback} image with control zones and human pose estimation  \label{element:3}
    \item \textbf{information pane} contains information about current UAV height and linear velocity \label{element:4}
    \item \textbf{video stream} from UAV camera
\end{enumerate}

\begin{figure}[h]
\centering
\includegraphics[width=1.0\columnwidth]{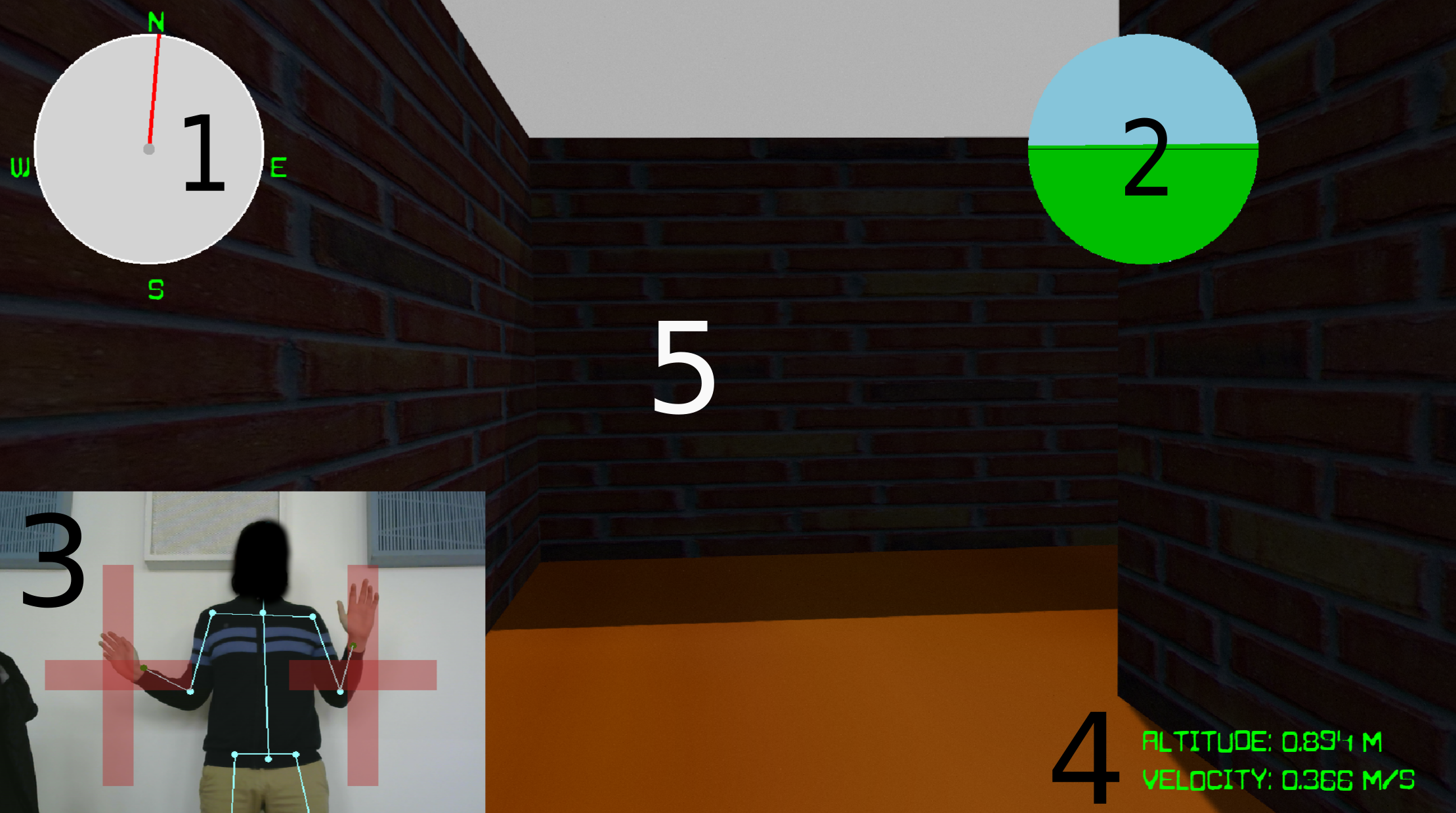}
\caption{First person view perspective which operator sees when operating UAV in simulation environment}
\label{fig:fpv}
\end{figure}

Simulated UAV odometry is used to extract neccessary information for elements \ref{element:1}, \ref{element:2} and \ref{element:3}. Visual feedback is used to enable FPV for UAV operator. What operator sees is shown in Figure \ref{fig:fpv}. When using RC joystick control, element \ref{element:3}) is excluded from FPV. 

% Razmisliti da li pisati formule za compass i artificial horizon kada su to elementi koji vec postoje 
% Svakako staviti sliku tog UI-a 
%

\section{Experimental setup}
\label{section:Experimental setup}

\noindent

Main goal of our experiment is to compare different control modalities for UAV and evaluate task load. Used controlled modalities are RC control with joystick and human pose control. To evaluate different control modalities, simulation environment and specifically designed task were used to compare operator performance. 

\subsection{System architecture}

The system architecture is composed of the hardware architecture and the software architecture. The hardware architecture refers to the hardware used for experiments. The software architecture refers to the software components used to develop and evaluate human pose control.

\subsubsection{Hardware architecture}

Simulation environment and experiments were done on PC. Server PC configuration is Intel Core i7-10700 CPU @ 2.9 GHz x 16, an Nvidia GeForce RTX 3090 24 GB GPU, and 64 GB RAM.
RaspberryPi 4 Model B was used as client PC for streaming visual feedback. 
Logitech C270 HD webcam was used to capture human pose. 
Futaba 10J RC controller was used as joystick. 
Arduino Nano was used to translate radio signals to ROS messages.
Epson Moverio BT-300 was used as operator FPV glasses.

\subsubsection{Software architecture}

To ensure experiment repeatability, GAZEBO 11 simulation environment on Ubuntu 20.04 was used. Model of the Parrot Bebop 2 UAV in GAZEBO was used. ROS Melodic was used to implement human pose control, RC control, UAV control and logging of experiments. All ROS nodes and simulation were executed on Server PC. 

For low-latency (less than 30 ms) video streaming RTSP protocol, GStreamer and Ubuntu Mate 20.04 on Raspberry Pi 4  were used. 
% Code is made available at: {link}

\begin{figure}[htbp]
\centerline{\includegraphics[width=1.0\columnwidth]{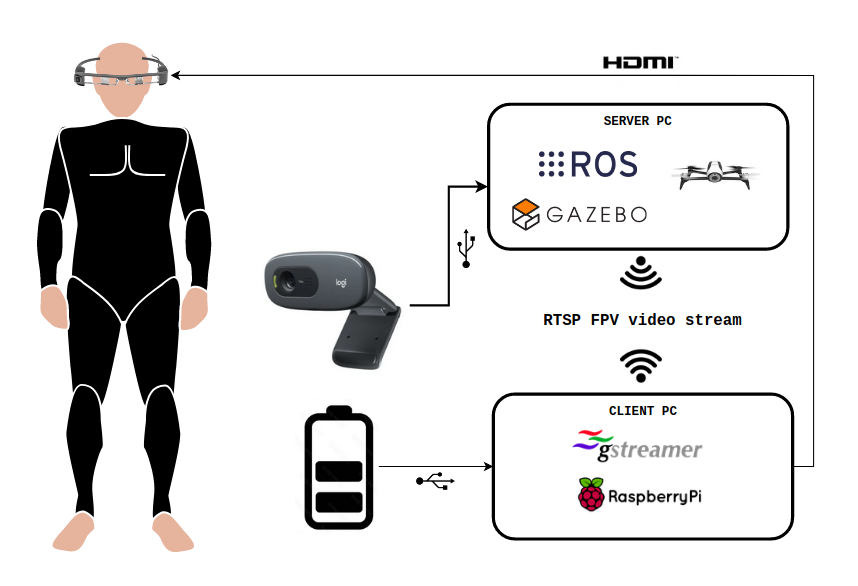}}
\caption{Experimental setup that shows how camera, googles and PCs were setup to achieve low-latency FPV stream to enable realtime control}
\label{fig:setup}
\end{figure}

In Figure \ref{fig:setup} it is shown how different PCs were setup with peripherals in order to achieve low latency RTSP streaming which enable realtime control. 

\subsection{RC controller}

To enable translation of radio signals from RC reciever to ROS message Arduino Nano was used. With \texttt{rosserial arduino}\footnote{\url{http://wiki.ros.org/rosserial_arduino/}}, PWM outputs from different channels of RC controller were transformed to \texttt{sensor\_msgs/Joy} ROS message which is used for joystick inputs.
% TODO: Maybe add code? 

% Make better name for it :) 
\subsection{UAV control architecture}

For UAV control, cascaded PID control for attitude and height was used. Software implementation of used controllers is presented in  \cite{Arbanas2018}. In Table \ref{tab:PID} PID gains are presented. The PID gains were tuned experimentally to achieve controllable UAV dynamics. 

\begin{table}[]
\centering
\caption{UAV control PID parameters}
\label{tab:PID}
\begin{tabular}{|l|c|c|c|l}
\cline{1-4}
\multicolumn{1}{|c|}{} & P   & I   & D   &  \\ \cline{1-4}
Roll                   & 10 & 0.25 & 0.25 &  \\ \cline{1-4}
Roll rate              & 50 & 50 & 0 &  \\ \cline{1-4}
Pitch                  & 10 & 0.25 & 0.25 &  \\ \cline{1-4}
Pitch rate             & 50 & 50 & 0 &  \\ \cline{1-4}
Yaw                    & 2.5 & 1.0 & 0.1 &  \\ \cline{1-4}
Yaw rate               & 30 & 0.0 & 0.0 &  \\ \cline{1-4}
z                      & 0.5 & 0.125 & 0.0 &  \\ \cline{1-4}
z rate                 & 75 & 10.0 & 0.41 &  \\ \cline{1-4}
\end{tabular}
\end{table} In Table \ref{tab:Scaling_factors} scaling factors are presented. Scaling factors are experimentally introduced to enable easy UAV control for participants.   

\begin{table}[h]
\centering
\caption{ Scaling factors for UAV control}
\label{tab:Scaling_factors}
\begin{tabular}{|c|c|c|c|c|}
\hline
               & z & $\phi$ & $\theta$ & $\psi$  \\ \hline
$s$ & 0.01   & 0.15 & 0.15  & 0.06 \\ \hline
\end{tabular}
\end{table}

Attitude control of UAV is achieved as follows: 

\begin{equation}
    \phi[k] = r_{4}s_{\phi}
    \label{eq:phi}
\end{equation}
\begin{equation}
    \theta[k] = r_{3}s_{\theta}
    \label{eq:theta}
\end{equation}
\begin{equation}
    \psi[k] = r_{2}s_{\psi} + \psi[k-1]
    \label{eq:psi}
\end{equation}
\begin{equation}
    z[k] = r_{1}s_{z} + z[k-1]
    \label{eq:z}
\end{equation}

% Add measurement units maybe? 

\subsection{Simulation setup}

\begin{figure}[htbp]
\centerline{\includegraphics[width=1.0\columnwidth]{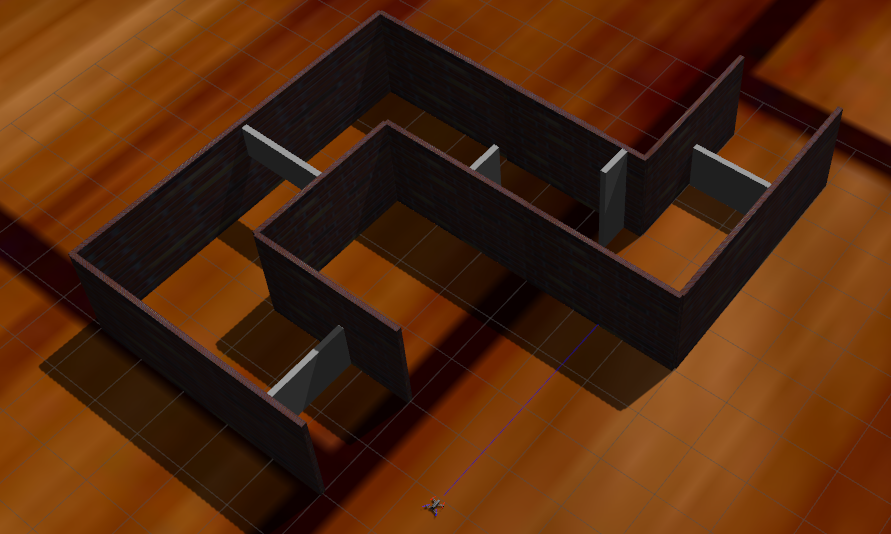}}
\caption{Maze in GAZEBO environment which was used for participants task load evaluation}
\label{fig:maze}
\end{figure}

To test the proposed control modalities, a simulated maze, shown in Figure \ref{fig:maze}, was created. The operator's task was to traverse the maze using FPV in combination with RC or HPE control. The order of control methods used depended on the experimental group.
 
\subsection{Experimental procedure}

For the experimental evaluation, 12 participants had to traverse the maze with both control modalities in a simulation with FPV. A between-participant experimental design was used. The first group of six participants traversed the maze first with the RC control and then with the HPE control. The second group of six participants traversed the maze first with the HPE control and then with the RC control. Before the runs, participants were given a questionnaire to complete. The questionnaire can be found in Table \ref{tab:questionnare_participants}. After each run, participants were given Raw NASA TLX, to assess workload, and a blank sheet of paper for qualitative feedback on the control modality used. For each maze traversion, we recorded the commanded references, the UAV odometry and traversion time.

\section{Experimental results}
\label{section:Experiments}

%https://humanrobotinteraction.org/2022/full-papers/

\subsection{NASA Task Load Index}

The NASA TLX is a multi-dimensional rating procedure that provides an overall workload score based on a weighted average of ratings on six subscales: Mental Demand, Physical Demand, Temporal Demand, Performance, Effort, and Frustration. The test consists of two parts\cite{NASAtlx}. The first part is used to determine the subjective importance (e.g. weight) of each subscale that contributes to the overall workload. The second part is used to rate from 0-20 (low to high) every demand, effort and frustration, and performance (good to bad ) from 0-20 (low to high) as they all contribute to overall workload.

The first part of the NASA TLX ratings is omitted. It makes sense to use adjusted ratings when scale is robust to beta change. Beta change occurs when participants re-anchor their ratings according to the nature of the task\cite{Bustamante2008}.
% Available here: https://journals.sagepub.com/doi/pdf/10.1177/154193120805201946
Since the nature of the task changes (different control modalities), we cannot guarantee robustness to beta change. Therefore, it is reasonable to use raw TLX (RTLX) ratings\cite{Hart2006}. 
% Available here: https://journals.sagepub.com/doi/pdf/10.1177/154193120605000909

% Two parts
% First part -> subjective importance of each category
% Second part -> rating of each of the scales 

\subsection{Population sample}

Experiments were conducted in collaboration with 12 participants. Participants ranged in age from 22-40 (\textit{mean}: 26.66, \textit{std}: 5.01) and were predominantly male (11/12).

% Add mean deviation (average age - std deviation) 

A limitation of the study arises from small diversity of the study population, which consisted mainly of young male university students. It is not known to what extent age and gender influence the subjective experience of workload. Such discrepancies might reveal interesting differences between age and gender groups rather than invalidate the proposed control technique. 
%Another limitation comes from the small diversity of our study population, which consisted mainly of young, male university students. It is unknown to which extent experience and observation shape the human representation of noninnate behaviors such as flight. We can therefore not exclude that factors such as age, gender, physical condition, or familiarity with technology could lead to the identification of different body motion pat-terns. However, such discrepancies may highlight interesting causes in motor learning and representation rather than invalidating the proposed identification method

\subsection{Results and Discussion}

\begin{figure}[htbp]
\centerline{\includegraphics[width=1.0\columnwidth]{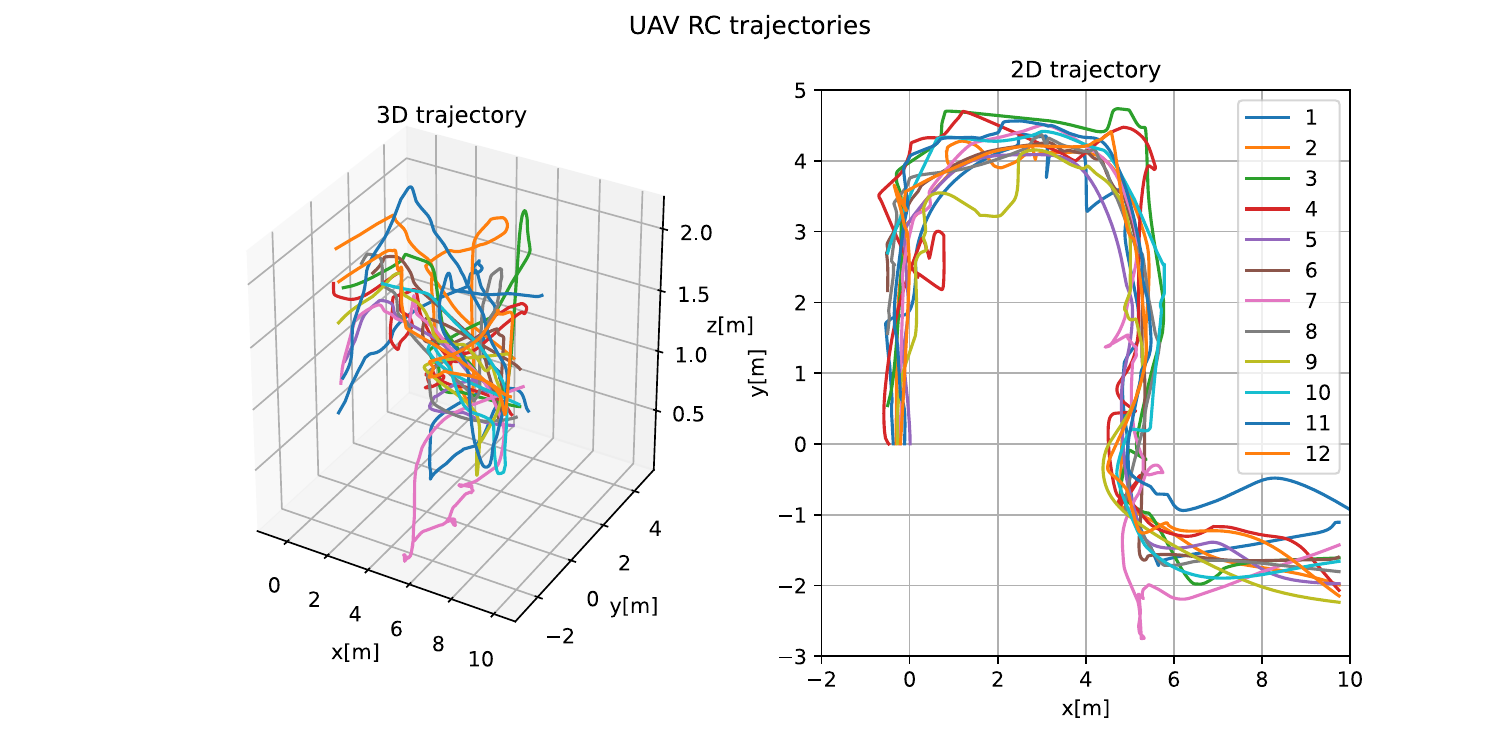}}
\caption{Trajectories participants achieved with RC control shown in 3D space and in x-y plane}
\label{fig:rc_trajectories}
\end{figure}

In Figure \ref{fig:rc_trajectories},
UAV trajectories participants achieved with RC control are shown. 

\begin{figure}[htbp]
\centerline{\includegraphics[width=1.0\columnwidth]{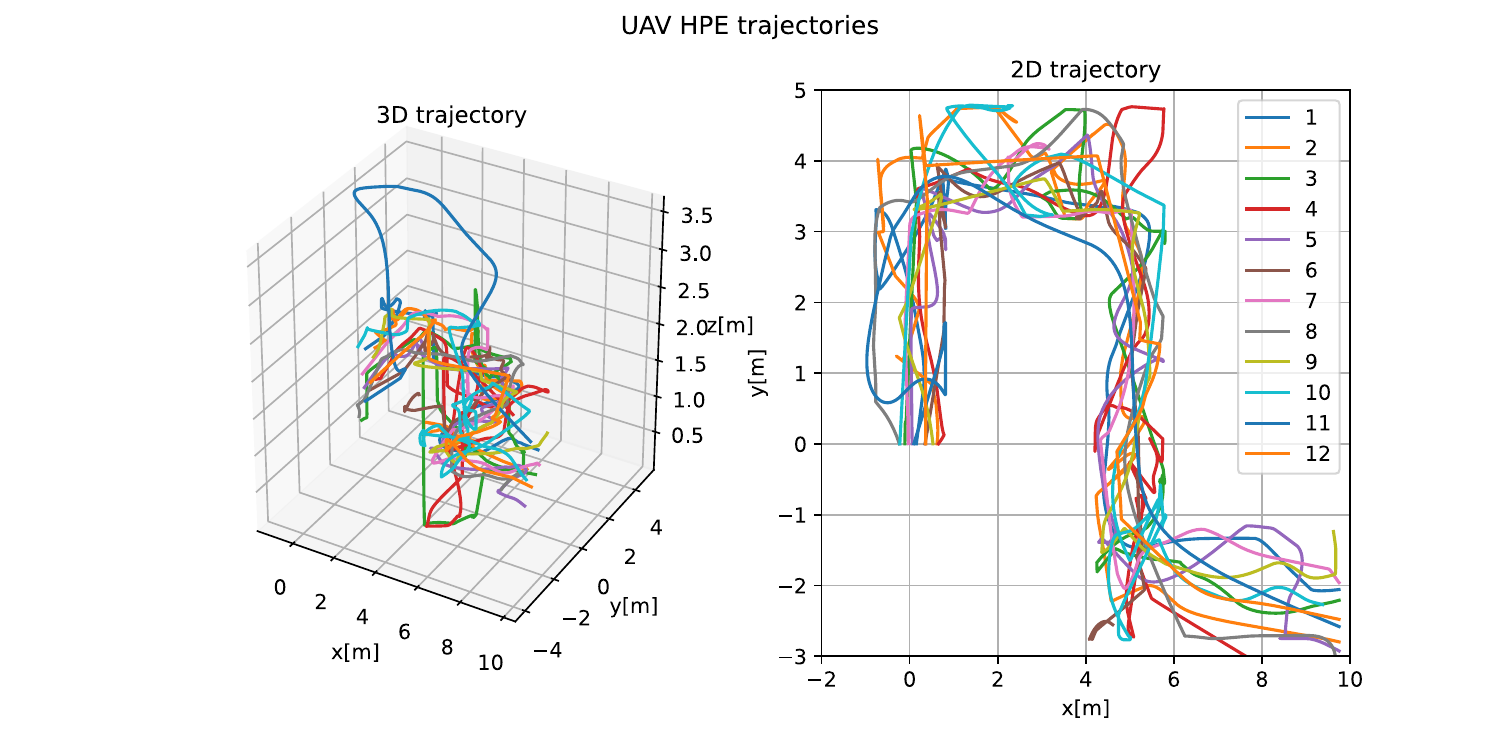}}
\caption{Trajectories participants achieved with HPE control shown in 3D space and in x-y plane}
\label{fig:hpe_trajectories}
\end{figure}

In Figure \ref{fig:hpe_trajectories}, UAV trajectories participants achieved with HPE control are shown. 

\begin{figure}[htbp]
\centerline{\includegraphics[width=1.0\columnwidth]{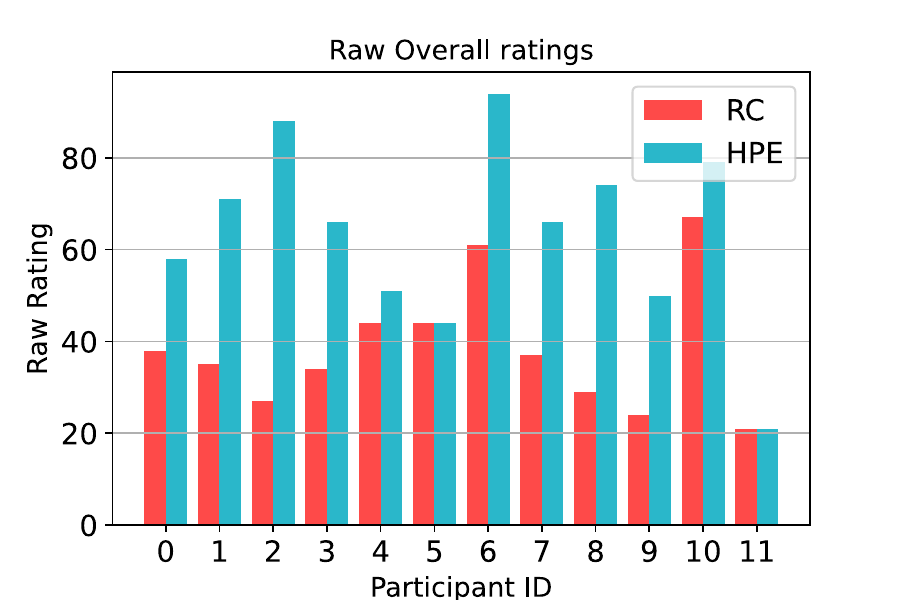}}
\caption{Overall raw NASA TLX score for each participant}
\label{fig:nasa_overall}
\end{figure}

\begin{figure}[htbp]
\centerline{\includegraphics[width=1.0\columnwidth]{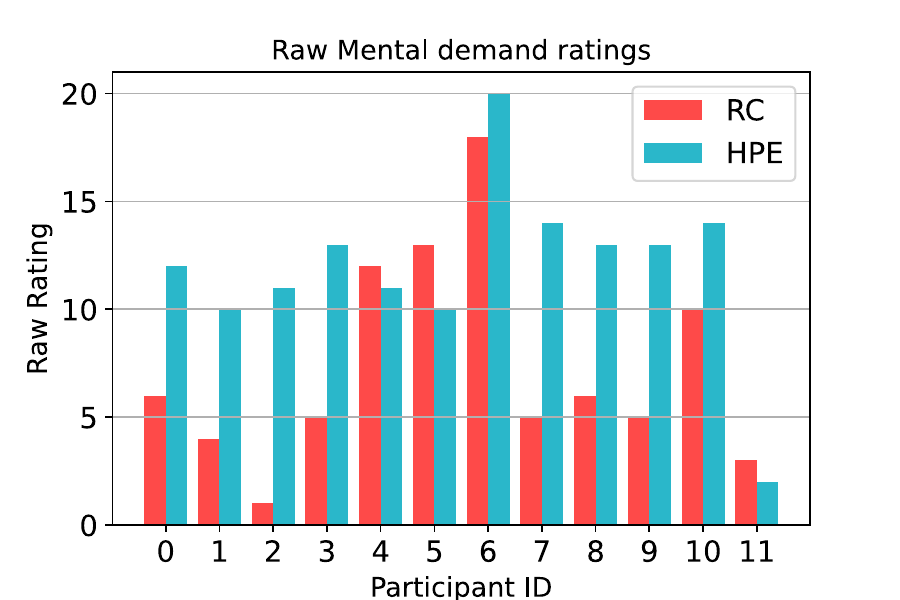}}
\caption{Mental demand raw NASA TLX score for each participant (lower is better)}
\label{fig:mental_demand}
\end{figure}

\begin{figure}[htbp]
\centerline{\includegraphics[width=1.0\columnwidth]{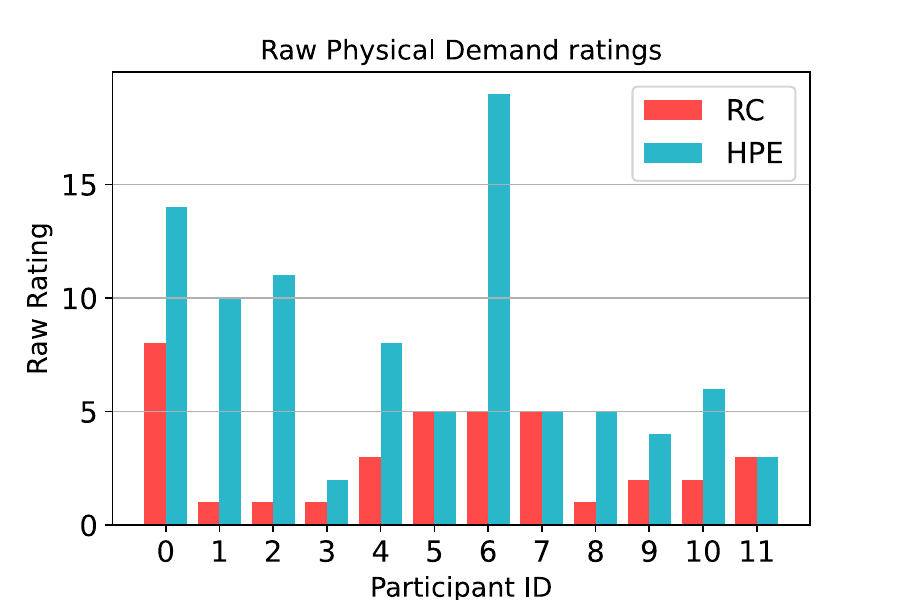}}
\caption{Physical demand raw NASA TLX score for each participant (lower is better)}
\label{fig:physical_demand}
\end{figure}

\begin{figure}[htbp]
\centerline{\includegraphics[width=1.0\columnwidth]{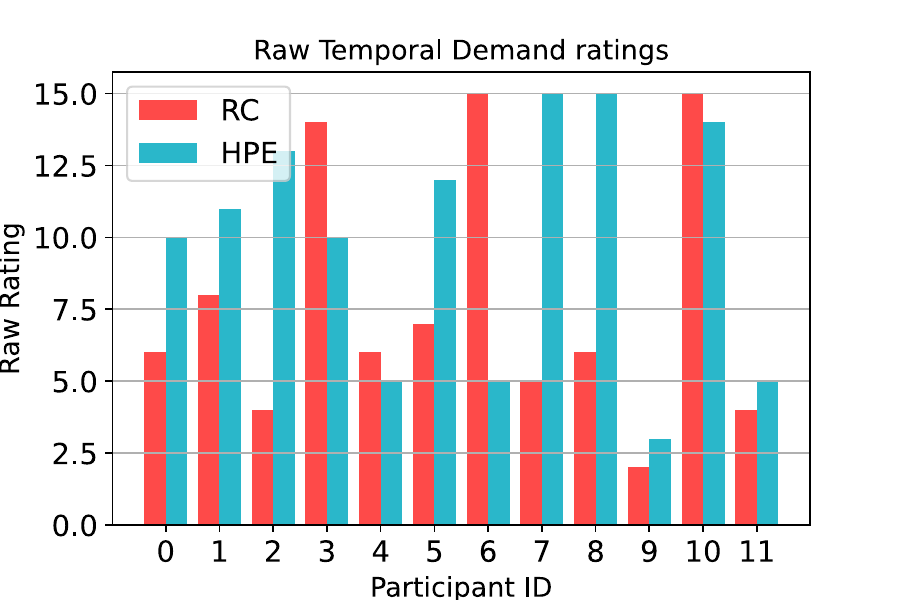}}
\caption{Temporal demand raw NASA TLX score for each participant (lower is better)}
\label{fig:temporal_demand}
\end{figure}

\begin{figure}[htbp]
\centerline{\includegraphics[width=1.0\columnwidth]{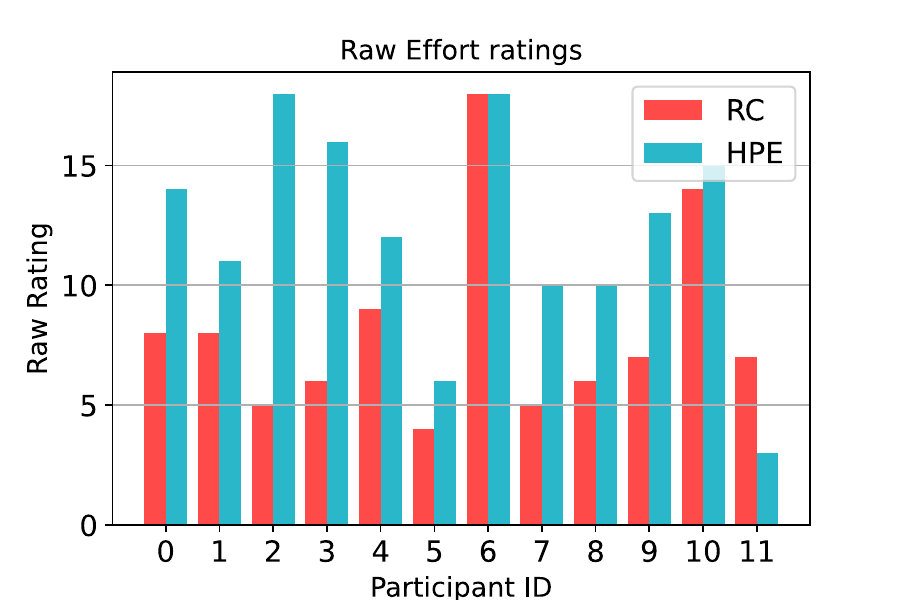}}
\caption{Effort raw NASA TLX score for each participant (lower is better)}
\label{fig:effrt}
\end{figure}

\begin{figure}[htbp]
\centerline{\includegraphics[width=1.0\columnwidth]{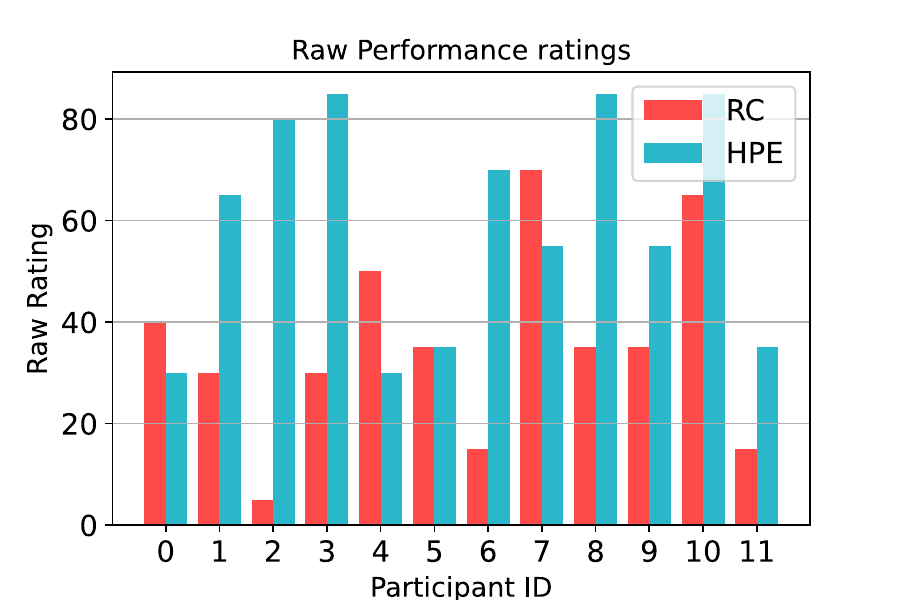}}
\caption{Performance raw NASA TLX score for each participant (lower is better)}
\label{fig:performance}
\end{figure}

\begin{figure}[htbp]
\centerline{\includegraphics[width=1.0\columnwidth]{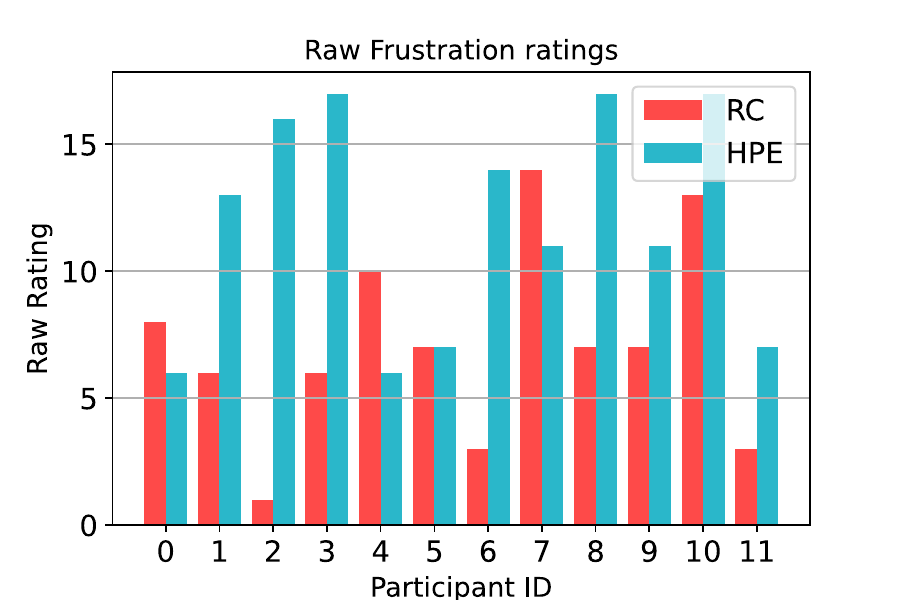}}
\caption{Frustration raw NASA TLX score for each participant (lower is better)}
\label{fig:frustration}
\end{figure}

The results of the overall raw data NASA TLX for each participant are shown in Figure \ref{fig:nasa_overall}. Each participant had a higher workload with the HPE control compared to the RC control. In its current form, the HPE control is more difficult to use than the RC control. In addition to the overall score, for clarity, the other RTLX subscales are also presented in Section \ref{section:appendix} with Figures \ref{fig:mental_demand}, \ref{fig:physical_demand},
\ref{fig:temporal_demand}, \ref{fig:effrt},  \ref{fig:performance}, \ref{fig:frustration}.

\begin{figure}[htbp]
\centerline{\includegraphics[width=0.8\columnwidth]{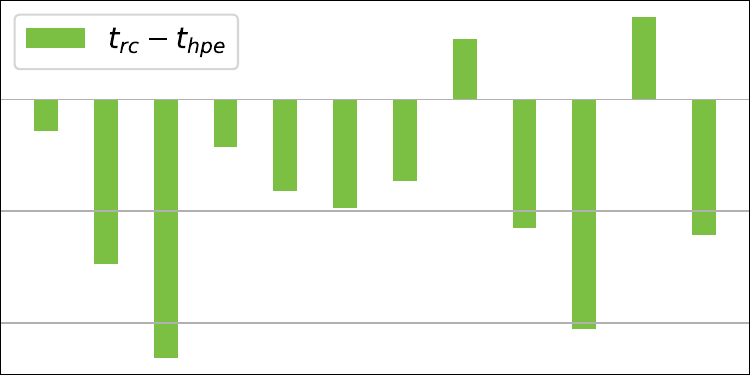}}
\caption{Traversion time difference $t_{rc} - t_{hpe}$ for each participant}
\label{fig:time_difference}
\end{figure}

Figure \ref{fig:time_difference}. shows the time differences between the HPE control and the RC control. Based on the results, in current form, almost all participants (10/12) traversed the maze faster with the RC control than with the HPE control. Interestingly, two participants who firstly used the HPE control were faster with it than with the RC control. These results can be attributed to familiarity with joystick control. Next to keyboard and mouse, joystick controllers are the most common interface when playing video games. Most participants had prior experience of UAV control as well as video games, as shown in Section \ref{section:appendix} in Table \ref{tab:questionnare_participants}.

\begin{figure}[htbp]
\centerline{\includegraphics[width=0.8\columnwidth]{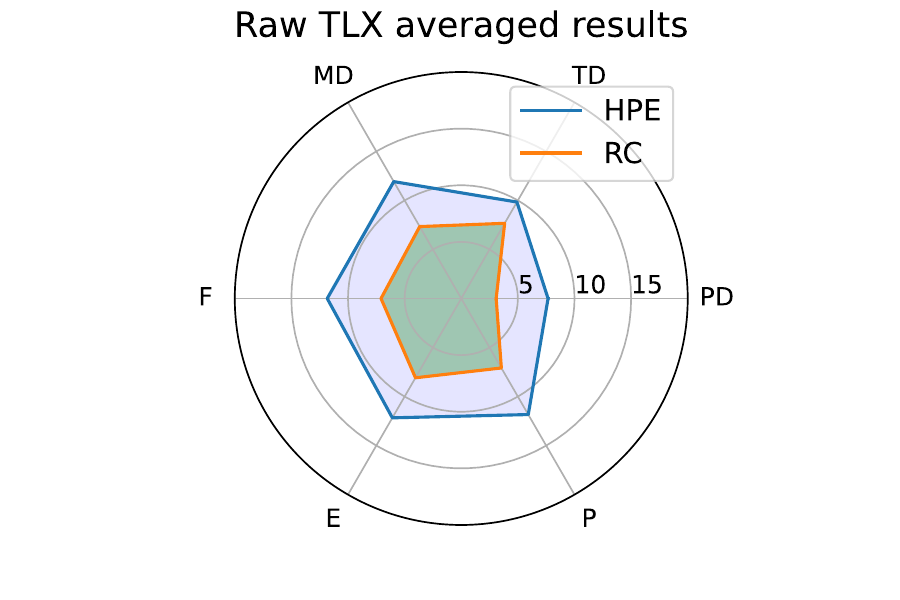}}
\caption{Averaged raw NASA TLX subscales of different control modalites for each participant}
\label{fig:spider_web}
\end{figure}

In Figure \ref{fig:spider_web}, the raw NASA TLX subscales are averaged for each participant. In its current form, the HPE control is worse than the RC control in every aspect that contributes to overall workload.

\subsection{Qualitative results}

After each run, participants were given a blank sheet of paper on which to record their observations of the control experiences. There are few observations shared by the experimental participants that could improve the HPE control:
\begin{enumerate}
    \item Adding depth could make pitch control more intuitive.
    \item The right hand should control yaw and pitch, roll control is redundant. 
    \item Yaw should be controlled with head movements (natural environment exploration).
    \item Constantly looking at my pose and the maze, diverted a lot of my attention and affected my performance.
    \item Virtual reality would make FPV control easier.
\end{enumerate}

% Main hypothesis is that intuitive control must be easy. Interesting thing

% Plotovi: 

% Position trajectory for sure 

% I have plots with average filter which is valuable but inco

\section{Conclusions}

In this paper we present a novel control modality based on continuous body pose gestures recognition with visual feedback. 
We conducted an experimental evaluation of the developed control modality compared to the most commonly used RC joystick in a simulation environment with 12 participants. The experimental evaluation included comparison of task execution (maze traversion) with different control modalities, questionnaire, raw NASA TLX and qualitative feedback.
We can conclude that human pose control in its current form is not yet developed to the point where it is easier to use in terms of task load. However, some of the results show promising potential, especially considering that each operator was using this control modality for the first time. 

Future work will focus on improving human pose control by incorporating depth and a different configuration of the interface. We also plan to use this control method for aerial manipulations. The qualitative feedback collected will be used to improve the control modality developed.

\label{section:Conclusions}
\section*{acknowledgement}

This work is supported by the project AERIAL COgnitive Integrated Multi-task Robotic System with Extended Operation Range and Safety (AERIAL-CORE) EU-H2020-ICT (grant agreement No. 871479).
We would like to thank everyone who devoted their time and energy to participate in experiments. 
%%%% Reference
\bibliographystyle{IEEEtran}
\bibliography{my_bibliography}

\begin{thebibliography}{10}
\providecommand{\url}[1]{#1}
\csname url@samestyle\endcsname
\providecommand{\newblock}{\relax}
\providecommand{\bibinfo}[2]{#2}
\providecommand{\BIBentrySTDinterwordspacing}{\spaceskip=0pt\relax}
\providecommand{\BIBentryALTinterwordstretchfactor}{4}
\providecommand{\BIBentryALTinterwordspacing}{\spaceskip=\fontdimen2\font plus
\BIBentryALTinterwordstretchfactor\fontdimen3\font minus \fontdimen4\font\relax}
\providecommand{\BIBforeignlanguage}[2]{{%
\expandafter\ifx\csname l@#1\endcsname\relax
\typeout{** WARNING: IEEEtran.bst: No hyphenation pattern has been}%
\typeout{** loaded for the language `#1'. Using the pattern for}%
\typeout{** the default language instead.}%
\else
\language=\csname l@#1\endcsname
\fi
#2}}
\providecommand{\BIBdecl}{\relax}
\BIBdecl

\bibitem{Agbeyangi2016}
A.~O. Agbeyangi, J.~O. Odiete, and A.~B. Olorunlomerue, ``{Review on UAVs used for Aerial Surveillance},'' \emph{Journal of Multidisciplinary Engineering Science and Technology}, vol.~3, no.~10, pp. 5713--5719, 2016.

\bibitem{Ren2019}
\BIBentryALTinterwordspacing
H.~Ren, Y.~Zhao, W.~Xiao, and Z.~Hu, ``{A review of UAV monitoring in mining areas: current status and future perspectives},'' \emph{International Journal of Coal Science and Technology}, vol.~6, no.~3, pp. 320--333, 2019. [Online]. Available: \url{https://doi.org/10.1007/s40789-019-00264-5}
\BIBentrySTDinterwordspacing

\bibitem{Butila2022}
E.~V. Butilă and R.~G. Boboc, ``{Urban Traffic Monitoring and Analysis Using Unmanned Aerial Vehicles (UAVs): A Systematic Literature Review},'' \emph{Remote Sensing}, vol.~14, no.~3, 2022.

\bibitem{Li2021}
X.~Li, Z.~Li, H.~Wang, and W.~Li, ``{Unmanned Aerial Vehicle for Transmission Line Inspection: Status, Standardization, and Perspectives},'' \emph{Frontiers in Energy Research}, vol.~9, no. July, pp. 1--13, 2021.

\bibitem{Jordan2018}
\BIBentryALTinterwordspacing
S.~Jordan, J.~Moore, S.~Hovet, J.~Box, J.~Perry, K.~Kirsche, D.~Lewis, and Z.~T.~H. Tse, ``{State-of-the-art technologies for UAV inspections},'' \emph{IET Radar, Sonar \& Navigation}, vol.~12, no.~2, pp. 151--164, 2018. [Online]. Available: \url{https://ietresearch.onlinelibrary.wiley.com/doi/abs/10.1049/iet-rsn.2017.0251}
\BIBentrySTDinterwordspacing

\bibitem{Herdel2022}
V.~Herdel, L.~Yamin, and J.~Cauchard, ``\BIBforeignlanguage{English}{Above and beyond: A scoping review of domains and applications for human-drone interaction},'' in \emph{\BIBforeignlanguage{English}{CHI 2022 - Proceedings of the 2022 CHI Conference on Human Factors in Computing Systems}}.\hskip 1em plus 0.5em minus 0.4em\relax Association for Computing Machinery, 2022.

\bibitem{Kellerman2020}
\BIBentryALTinterwordspacing
R.~Kellermann, T.~Biehle, and L.~Fischer, ``{Drones for parcel and passenger transportation: A literature review},'' \emph{Transportation Research Interdisciplinary Perspectives}, vol.~4, p. 100088, 2020. [Online]. Available: \url{https://www.sciencedirect.com/science/article/pii/S2590198219300879}
\BIBentrySTDinterwordspacing

\bibitem{Kratky2021}
V.~Krátký, A.~Alcántara, J.~Capitán, P.~Štěpán, M.~Saska, and A.~Ollero, ``Autonomous aerial filming with distributed lighting by a team of unmanned aerial vehicles,'' \emph{IEEE Robotics and Automation Letters}, vol.~6, no.~4, pp. 7580--7587, 2021.

\bibitem{Funk2018}
\BIBentryALTinterwordspacing
M.~Funk, ``Human-drone interaction: Let's get ready for flying user interfaces!'' \emph{Interactions}, vol.~25, no.~3, p. 78–81, apr 2018. [Online]. Available: \url{https://doi.org/10.1145/3194317}
\BIBentrySTDinterwordspacing

\bibitem{Stiefelhagen2007}
R.~Stiefelhagen, H.~K. Ekenel, C.~Fugen, P.~Gieselmann, H.~Holzapfel, F.~Kraft, K.~Nickel, M.~Voit, and A.~Waibel, ``Enabling multimodal human–robot interaction for the karlsruhe humanoid robot,'' \emph{IEEE Transactions on Robotics}, vol.~23, no.~5, pp. 840--851, 2007.

\bibitem{Cherpillod2019}
A.~Cherpillod, D.~Floreano, and S.~Mintchev, ``Embodied flight with a drone.''\hskip 1em plus 0.5em minus 0.4em\relax Institute of Electrical and Electronics Engineers Inc., 2019, pp. 386--390.

\bibitem{NASAtlx}
\BIBentryALTinterwordspacing
H.~S. G., ``Nasa task load index (tlx),'' in \emph{NASA Ames Research Center Moffett Field, CA United States)}, January 1986. [Online]. Available: \url{https://ntrs.nasa.gov/citations/20000021488}
\BIBentrySTDinterwordspacing

\bibitem{Marinov2021}
\BIBentryALTinterwordspacing
Z.~Marinov, S.~Vasileva, Q.~Wang, C.~Seibold, J.~Zhang, and R.~Stiefelhagen, ``Pose2drone: A skeleton-pose-based framework for human-drone interaction,'' 2021. [Online]. Available: \url{http://arxiv.org/abs/2105.13204}
\BIBentrySTDinterwordspacing

\bibitem{Cao2018}
\BIBentryALTinterwordspacing
Z.~Cao, G.~Hidalgo, T.~Simon, S.-E. Wei, and Y.~Sheikh, ``Openpose: Realtime multi-person 2d pose estimation using part affinity fields,'' 2018. [Online]. Available: \url{http://arxiv.org/abs/1812.08008}
\BIBentrySTDinterwordspacing

\bibitem{Tezza2019}
D.~Tezza and M.~Andujar, ``The state-of-the-art of human-drone interaction: A survey,'' \emph{IEEE Access}, vol.~7, pp. 167\,438--167\,454, 2019.

\bibitem{Fernandez2016}
R.~A. Fernandez, J.~L. Sanchez-Lopez, C.~Sampedro, H.~Bavle, M.~Molina, and P.~Campoy, ``Natural user interfaces for human-drone multi-modal interaction.''\hskip 1em plus 0.5em minus 0.4em\relax Institute of Electrical and Electronics Engineers Inc., 2016, pp. 1013--1022.

\bibitem{Macchini2020}
\BIBentryALTinterwordspacing
M.~Macchini, J.~Frogg, F.~Schiano, and D.~Floreano, ``Does spontaneous motion lead to intuitive body-machine interfaces? a fitness study of different body segments for wearable telerobotics,'' 2020. [Online]. Available: \url{http://arxiv.org/abs/2011.07591}
\BIBentrySTDinterwordspacing

\bibitem{Zhao}
J.~Zhao, R.~An, R.~Xu, and B.~Lin, ``{Comparing Hand Gestures and a Gamepad Interface for Locomotion in Virtual Environments},'' pp. 1--14.

\bibitem{Pfeil2013}
K.~P. Pfeil, S.~L. Koh, and J.~J. LaViola, ``{Exploring 3D gesture metaphors for interaction with unmanned aerial vehicles},'' \emph{International Conference on Intelligent User Interfaces, Proceedings IUI}, pp. 257--266, 2013.

\bibitem{Miehlbradt2018}
\BIBentryALTinterwordspacing
J.~Miehlbradt, A.~Cherpillod, S.~Mintchev, M.~Coscia, F.~Artoni, D.~Floreano, and S.~Micera, ``{Data-driven body\&\#x2013;machine interface for the accurate control of drones},'' \emph{Proceedings of the National Academy of Sciences}, vol. 115, no.~31, pp. 7913--7918, 2018. [Online]. Available: \url{https://www.pnas.org/doi/abs/10.1073/pnas.1718648115}
\BIBentrySTDinterwordspacing

\bibitem{Chen2019}
\BIBentryALTinterwordspacing
L.~Chen, A.~Ebi, K.~Takashima, K.~Fujita, and Y.~Kitamura, ``Pinpointfly: An egocentric position-pointing drone interface using mobile ar,'' in \emph{SIGGRAPH Asia 2019 Emerging Technologies}, ser. SA '19.\hskip 1em plus 0.5em minus 0.4em\relax New York, NY, USA: Association for Computing Machinery, 2019, p. 34–35. [Online]. Available: \url{https://doi.org/10.1145/3355049.3360534}
\BIBentrySTDinterwordspacing

\bibitem{xiao2018simple}
B.~Xiao, H.~Wu, and Y.~Wei, ``Simple baselines for human pose estimation and tracking,'' 2018.

\bibitem{andriluka14cvpr}
M.~Andriluka, L.~Pishchulin, P.~Gehler, and B.~Schiele, ``2d human pose estimation: New benchmark and state of the art analysis,'' in \emph{IEEE Conference on Computer Vision and Pattern Recognition (CVPR)}, 2014.

\bibitem{Arbanas2018}
\BIBentryALTinterwordspacing
B.~Arbanas, A.~Ivanovic, M.~Car, M.~Orsag, T.~Petrovic, and S.~Bogdan, ``Decentralized planning and control for uav--ugv cooperative teams,'' \emph{Autonomous Robots}, 2018. [Online]. Available: \url{https://doi.org/10.1007/s10514-018-9712-y}
\BIBentrySTDinterwordspacing

\bibitem{Bustamante2008}
\BIBentryALTinterwordspacing
E.~A. Bustamante and R.~D. Spain, ``{Measurement Invariance of the Nasa TLX},'' \emph{Proceedings of the Human Factors and Ergonomics Society Annual Meeting}, vol.~52, no.~19, pp. 1522--1526, 2008. [Online]. Available: \url{https://doi.org/10.1177/154193120805201946}
\BIBentrySTDinterwordspacing

\bibitem{Hart2006}
\BIBentryALTinterwordspacing
S.~G. Hart, ``{Nasa-Task Load Index (NASA-TLX); 20 Years Later},'' \emph{Proceedings of the Human Factors and Ergonomics Society Annual Meeting}, vol.~50, no.~9, pp. 904--908, 2006. [Online]. Available: \url{https://doi.org/10.1177/154193120605000909}
\BIBentrySTDinterwordspacing

\end{thebibliography}

% HOW TO ADD APPENDIX? 
\section{Appendix}
\label{section:appendix}

\begin{table*}[htbp]
\centering
\caption{Questionnare answers}
\label{tab:questionnare_participants}
\begin{tabular}{|c|c|c|c|c|c|c|c|}
\hline
User ID & Age & Gender & \begin{tabular}[c]{@{}c@{}}How many \\ times have you\\  operated UAV?\end{tabular} & \begin{tabular}[c]{@{}c@{}}Have you \\ ever been \\ professional \\ athlete?\end{tabular} & \begin{tabular}[c]{@{}c@{}}Have you ever\\  played \\ videogames\\  on your PC?\end{tabular} & \begin{tabular}[c]{@{}c@{}}Have you \\ ever\\  had videogames \\ console?\end{tabular} & \begin{tabular}[c]{@{}c@{}}Have you ever \\ used virtual or \\ augumented\\  reality?\end{tabular} \\ \hline
1       & 27  & M      & 5 and more                                                                          & NO                                                                                        & YES                                                                                          & YES                                                                                    & YES                                                                                                \\ \hline
2       & 27  & M      & 1-3 times                                                                           & NO                                                                                        & YES                                                                                          & YES                                                                                    & NO                                                                                                 \\ \hline
3       & 25  & F      & 1-3 times                                                                           & NO                                                                                        & YES                                                                                          & NO                                                                                     & NO                                                                                                 \\ \hline
4       & 40  & M      & 3-5 times                                                                           & NO                                                                                        & YES                                                                                          & YES                                                                                    & NO                                                                                                 \\ \hline
5       & 26  & M      & 5 and more                                                                          & YES                                                                                       & YES                                                                                          & YES                                                                                    & YES                                                                                                \\ \hline
6       & 21  & M      & 1-3 times                                                                           & NO                                                                                        & YES                                                                                          & NO                                                                                     & NO                                                                                                 \\ \hline
7       & 29  & M      & Never                                                                               & NO                                                                                        & YES                                                                                          & NO                                                                                     & NO                                                                                                 \\ \hline
8       & 21  & M      & Never                                                                               & NO                                                                                        & YES                                                                                          & YES                                                                                    & NO                                                                                                 \\ \hline
9       & 28  & M      & 3-5 times                                                                           & NO                                                                                        & YES                                                                                          & YES                                                                                    & YES                                                                                                \\ \hline
10      & 26  & M      & 3-5 times                                                                           & NO                                                                                        & YES                                                                                          & NO                                                                                     & NO                                                                                                 \\ \hline
11      & 28  & M      & 1-3 times                                                                           & NO                                                                                        & YES                                                                                          & YES                                                                                    & NO                                                                                                 \\ \hline
12      & 22  & M      & Never                                                                               & YES                                                                                       & YES                                                                                          & YES                                                                                    & YES                                                                                                \\ \hline
\end{tabular}
\end{table*}

\end{document}